\documentstyle[12pt]{article}

\def\caption#1{{\centerline{\vbox{\baselineskip=15pt
       \vskip.15in\hsize=15cm\noindent{#1}\vskip.1mm }}}}
\renewcommand{\theequation}{\thesection.\arabic{equation}}
\newcommand{\be}{\begin{equation}}   \newcommand{\ee}{\end{equation}}
\newcommand{\bear}{\begin{eqnarray}}
\newcommand{\eear}{\end{eqnarray}}
\newcommand{\ba}{\begin{array}}      \newcommand{\ea}{\end{array}}

\newcommand{\up}{\mbox{$P$}}
\newcommand{\down}{\mbox{$N$}}
\newcommand{\inl}{{\scriptscriptstyle L}}
\newcommand{\inr}{{\scriptscriptstyle R}}

\begin{document}
\pagestyle{empty}
\begin{titlepage}
\def\thepage {}        % Kill page numbering

\title{{\bf Negative contributions to $S$\\
in an effective field theory}}

%\vspace*{2mm}
\author{Bogdan A. Dobrescu\thanks{Address after September 1, 1997:
Theoretical Physics Department, MS106, Fermi National Accelerator
Laboratory, P.O. Box 500, Batavia, IL 60510, USA.}\\
\\
{\small {\it Department of Physics}} \\
{\small {\it Boston University}}\\
{\small {\it 590 Commonwealth Avenue}}\\
{\small {\it Boston, MA 02215}} \\
\\
and\\
\\
John Terning\thanks{e-mail
  addresses: dobrescu@budoe.bu.edu,  terning@alvin.lbl.gov}\\
\\
{\small {\it Department of Physics}}\\
{\small {\it University of California}}\\
{\small {\it Berkeley, CA 94720}}\\ }

\date{ }

\maketitle

\vspace*{-155mm}

\noindent
\makebox[12.5cm][l]{BUHEP-97-27} September 19, 1997 \\
\makebox[12.5cm][l]{UCB-PTH-97/44} hep-ph/9709297 \\

\vspace*{140mm}

\baselineskip=18pt

\begin{abstract}

{\normalsize
We show that an effective field theory that includes non-standard
couplings between the electroweak gauge bosons and the top and
bottom quarks may yield negative contributions to both the $S$ and
$T$ oblique radiative electroweak parameters.  We find that that such
an effective field theory provides a better fit to data than the standard
model (the $\chi^2$ per degree of freedom is half as large). We
examine in some detail an illustrative model where the exchange of
heavy scalars produces the correct type of non-standard couplings.}

\end{abstract}

\vfill
\end{titlepage}

\baselineskip=18pt
\pagestyle{plain}
\setcounter{page}{1}

%%%%%%%%%%%%%%%%%%%%%%%%%%%%%%%%%%%%%%%%%%%%%
%%%%%%%%%%%%%%%%%%
\section{Introduction}

An important drawback of the standard model (SM) and its minimal
supersymmetric extensions is that there are no theoretical constraints
on the Yukawa couplings of the Higgs doublet, so that there are no
clues about the quark and lepton spectrum.
On the other hand, models which involve dynamical mechanisms for
fermion
mass generation, such as extended technicolor \cite{etc}, typically
induce corrections to precision electroweak observables that are in
disagreement
with data. Usually, the $S$ and $T$ oblique radiative electroweak
parameters
are too large \cite{ST}, and the coupling
of the $Z$ gauge boson to $b$ quarks is shifted so that the ratio
of $Z \rightarrow b{\bar b}$ to $Z \rightarrow$ hadrons branching
fractions is too small \cite{css}.

In this paper we show that, although shifts in oblique parameters and
weak gauge couplings may
individually be in disagreement with the precision electroweak data,
their combination may lead to a much better fit than the SM.
As we will demonstrate this possibility arises because shifts in weak
gauge couplings can produce significant contributions to the oblique
parameters.
In Section 2 we discuss an effective field theory in which
non-standard couplings of the electroweak gauge bosons to the third
generation quarks may yield negative contributions to both $S$ and
$T$. In Section 3 we fit this effective field theory to the
electroweak data. The couplings of the effective theory considered
here can be
produced, for example, by the exchange of heavy scalars in a
technicolor model, as we show in Section 4. Our conclusions are
presented in Section 5.

%%%%%%%%%%%%%%%%%%%%%%%%%%%%%%%%%%%%%%%%%%%%%
%%%%%%%%%%%
\section{An effective field theory calculation}
\setcounter{equation}{0}

In extensions of the SM, in addition to the mass terms for the
electroweak gauge bosons induced by the Higgs mechanism,
there are new terms which can be
generated in the effective Lagrangian below the scale of new physics.
Two\footnote{There is a third oblique correction parameter $U$, but it
is generally much smaller than $S$ and $T$ in models without extra
gauge bosons, and we will not consider it here.}
phenomenologically important terms (of dimension 2 and 4 respectively)
are \cite{ST,holdom}:
\be
{\cal L}^{\rm oblique}_{\rm eff} = - \alpha T^0 v^2 \, W^{3 \mu}
W^3_\mu
- {{S^0}\over{16 \pi}}  \, g g' \, B^{\mu \nu} W^3_{\mu \nu}
 ~,
\ee
where $W^3_{\mu \nu}$ and $B^{\mu \nu}$ are the gauge field strength
tensors corresponding to the neutral gauge bosons of the  $SU(2)_W
\times U(1)_Y$ group that mix to produce the physical $Z$ and the
photon,
$\alpha$ is the electromagnetic coupling constant, $g$ and $g^\prime$
are the weak and hypercharge gauge couplings, and $v$ is the weak
scale.
These terms give tree-level contributions to the oblique radiative
correction parameters
$S$ and $T$ which can be defined \cite{ST} as:
\be
S  \equiv  - \frac{8 \pi}{M_Z^2} \left( \Pi_{3Y}(M_Z^2)
- \Pi_{3Y}(0) \right) \label{expv} \\  ~,
\label{Sdef}
\ee
\be
T \equiv \frac{4}{\alpha v^2} \left[ \Pi_{11}(0) -
\Pi_{33}(0) \right] ~,
\ee
where $\Pi_{jk}(q^2)$  are the vacuum
polarizations of the $W^\mu_j$ and $B^\mu$ electroweak gauge fields
due to non-SM physics, with the gauge couplings factored
out\footnote{We are using the definition for hypercharge where
$Y \equiv 2(Q - T_3)$.}. Thus, in the effective theory we have
tree-level and loop contributions to the oblique parameters:
\be
S = S^0 + S^{\rm loop}~,
\ee
\be
T = T^0 + T^{\rm loop}~.
\ee

New physics in the electroweak symmetry breaking sector can also
induce changes in the interactions of quarks with electroweak gauge
bosons \cite{zhang,css,burgess}.
In dynamical models of electroweak symmetry breaking,
the new interactions are also responsible for generating masses for
quarks and leptons, so we would expect to see the largest effects in
the couplings of the top-bottom doublet.
Below the scale of the new physics we can parameterize
these effects with an effective Lagrangian which includes three new
parameters ($\delta g_L, \delta g_R^{t,b}$):
\bear
{\cal L}_{\rm eff}^{\rm vertex} & = & \delta g_L
\left(g W^\mu_j - g^\prime \delta_{3j} B^\mu \right)
\overline q_{\inl} \gamma_\mu  \sigma^j q_{\inl}
\nonumber \\ [2mm]
& & + \, \left(g W^\mu_3 - g^\prime B^\mu \right)
\left[\delta g_R^t (\overline{t}_{\inr}\gamma_{\mu} t_{\inr})
+ \delta g_R^b (\overline{b}_{\inr}\gamma_{\mu} b_{\inr}) \right]~,
\label{vertex}
\eear
where $q_{\inl} \equiv (t_{\inl}, b_{\inl})$ is the left-handed
$t-b$ quark doublet, and $\sigma_j$ are the Pauli matrices.
Such shifts in electroweak couplings produce tree-level effects in
precision electroweak observables.  In addition to these tree-level
contributions (which come
from integrating out physics above the effective field theory cutoff,
$\Lambda$) there are loop corrections \cite{loop} from physics below the scale
$\Lambda$ which renormalize the coefficients in
${\cal L}_{\rm eff} $.  If $\Lambda$ is significantly
larger than $M_Z$, then the leading logarithms from these one-loop
corrections may be numerically important.  For example in technicolor
models, light pseudo-Nambu-Goldstone
bosons can contribute to $S$ and $T$ at one-loop.
In this section we will calculate the one-loop contributions to $S$
and $T$ from the shifts in the $t$ and $b$ gauge couplings.

%%%%%%%%%%%%%%%%%%%%%%%%%%%%%%%%%%%%%%%%%%%%%
%%%%%%%%%%%%%%%%%%%%%%%%%%
\begin{figure}
\input FEYNMAN
\begin{picture}(20000,11500)(-10000,3000)
\THICKLINES
  \put(10700,10000){\oval(4150,4000)}
  \drawarrow[\E\ATTIP](10950,12000)
  \drawarrow[\W\ATTIP](10450,8000)
  \put(8000,10000){\circle*{1400}}
  \put(13400,10000){\circle*{1400}}
  \drawline\photon[\W\REG](8000,10000)[6]
  \drawline\photon[\E\REG](13200,10000)[6]

  \put(9900,12800){$t,b$}
% \put(13000,12000){$B$}
  \put(2500,11000){$W^3$}\put(18000,11000){$B$}

\end{picture}

\vspace{-1cm}

\caption{ {\small Contributions from $t$ and $b$  to $\Pi_{3Y}$. The
$\bullet$ represents the effective couplings of the quarks to the
gauge bosons.}}
\end{figure}
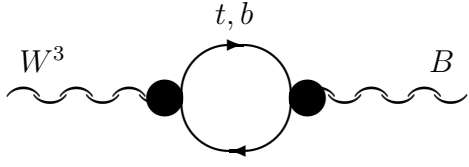
%%%%%%%%%%%%%%%%%%%%%%%%%%%%%%%%%%%%%%%%%%%%%
%%%%%%%%%%%%%%%%%%%%%%%%%%

It is convenient to use the following effective couplings:
\bear
\overline{g}_L^{t,b} &\equiv &  \pm \left(\frac{1}{2} + \delta g_L
\right)~,\nonumber \\
\overline{Y}^{t,b}_L &\equiv & Y_L \mp 2 \delta g_L ~,\nonumber \\
\overline{Y}^{t,b}_R  &\equiv & Y^{t,b}_R - 2 \delta g_R^{t,b}~,
\eear
where $Y_L = 1/3$, $Y^t_{R} = 4/3$, and $Y^b_{R}
= -2/3$ are the quark hypercharges.
The one-loop contribution of the $t$ quark to the vacuum polarization
(see Fig.~1) is then:
\bear
\Pi^{(t)}_{3Y}(q^2) &=& -\frac{N_c}{8 \pi^2} \int_0^1 d x
\left\{ 2  \left( \overline{g}_L^{t} \overline{Y}^t_{L}+ \delta g_R^t
\overline{Y}^t_{R} \right)
\left[ x(1 - x) q^2 -  \frac{m_t^2}{2} \right]
\right.\nonumber \\ [2mm]
& & + \left.
  \left( \overline{g}_L^{t} \overline{Y}^t_{R} + \delta g_R^t
\overline{Y}^t_{L} \right) m_t^2 \vphantom{\frac{m_t^2}{2}} \right\}
\ln\left[ \frac{\Lambda^2}{ m_t^2 - x(1-x) q^2 } \right]~,
\label{pol}
\eear
where $N_c = 3$, $m_t$ is the $t$ mass, and a similar
expression holds for the $b$ quark contribution.
Thus the one-loop result for the vacuum polarization is:
\bear
\Pi^{(t,b)}_{3Y}(M_Z^2) - \Pi^{(t,b)}_{3Y}(0) &\approx&
 -\frac{N_c}{12 \pi^2} M_Z^2
\left[ \left( \overline{g}_L^t \overline{Y}^t_{L} + \delta g_R^t
\overline{Y}^t_{R} \right)
\ln\left( \frac{\Lambda}{m_t} \right)
 \right.\nonumber \\ [2mm]
& & + \left.
\left( \overline{g}_L^b \overline{Y}^b_{L} +
\delta g_R^b \overline{Y}^b_{R}  \right)
\ln\left( \frac{\Lambda}{M_Z} \right) \right]~,
\label{dif}
\eear
where we kept only the leading-log terms (finite terms can be
absorbed into the definition of $S^0$ and $T^0$ through the
matching conditions at the scale $\Lambda$) and  ignored  terms
suppressed by $m_b^2/M_Z^2$. To obtain the contribution to $S$
we must subtract the SM contribution;
the result to leading order in  $ \delta g_{L,R}$ is
\bear
S^{(t,b)} & \approx & \frac{2 N_c}{3 \pi} \left\{ \left[
(Y_L - 1) \delta g_{L} +  Y^t_{R} \, \delta g_{R}^{t} \right]
\ln\left( \frac{\Lambda}{m_t} \right)
\right.\nonumber \\ [2mm]
& & +  \left.
\left[ - (Y_L + 1) \delta g_{L}
+ Y^b_{R} \, \delta g_{R}^{b}
\right]
\ln\left( \frac{\Lambda}{M_Z} \right) \right\} ~.
\eear
This result is quite general given a weak doublet of fermions with
masses
$m_b \ll M_Z < m_t$. Using the values of the hypercharges, and discarding
terms proportional to $\ln(m_t/M_Z)$, we get
\be
S^{(t,b)} \approx
- \frac{4}{3 \pi} \left( 3\, \delta g_L - 2\, \delta g_{R}^{t}
+ \delta g_{R}^{b} \right) \ln\left( \frac{\Lambda}{M_Z} \right) ~.
\ee
We see from this expression that the contribution to $S$ can be
of either sign (since the shifts in the couplings can be of either
sign) and large (it is enhanced with respect to the finite
contribution of a weak-doublet of fermions by the logarithm and by a
color factor).

We now move on to the isospin breaking effects which contribute to
$T$. There are no corrections to $T$ from $\delta g_L$;
the only large correction to $T$ is due to $\delta g^t_{R}$
(a similar effect is discussed in \cite{cdt}):
\be
\Pi_{11}^{(t,b)}(0) - \Pi_{33}^{(t,b)}(0) \approx  \delta g_{R}^{t} \,
\frac{N_c}{4 \pi^2} \,
m_t^2 \, \ln\left( \frac{\Lambda}{m_t} \right)~,
\ee
which gives
\be
T^{(t,b)} \approx \delta g_{R}^{t} \frac{3 m_t^2}{\pi^2\alpha v^2}
\ln\left( \frac{\Lambda}{m_t} \right)~.
\label{direct}
\ee
Again we see that the contribution can be of either sign.  We also
note that $\delta g_{R}^{t}  < 0$ induces negative contributions for
both $S$ and $T$.

%%%%%%%%%%%%%%%%%%%%%%%%%%%%
%%%%%%%%%%%%%%%%%%%%%%%%%%%%
\section{Comparison with Experiment}
\setcounter{equation}{0}

In order to assess the usefulness of this formalism we have performed
a fit to the precision electroweak data using the standard techniques
\cite{burgess,bufits} with the parameters $S$, $T$, $\delta g_L$, and
$\delta g_R^b$ (at present there is no direct precision measurement
involving $\delta g_R^t$). The deviations from the
SM predictions for the physical quantities used in the fit in terms of
these parameters are given in the Appendix.
The experimental values \cite{data,data1} and SM
predictions \cite{langacker} as well as the best fit values are
given in Table 1.

\begin{table}[htbp]
\begin{center}
\begin{tabular}{|c|l|l|l|}\hline\hline
Quantity & Experiment & SM & Fit \\\hline
$\Gamma_Z$ & 2.4947 $\pm$ 0.0026 & 2.4925 & 2.4948  \\
$R_e$ & 20.756 $\pm$ 0.057 & 20.717 & 20.789  \\
$R_\mu$ & 20.795 $\pm$ 0.039 & 20.717 & 20.789  \\
$R_\tau$ & 20.831 $\pm$ 0.054 & 20.717 & 20.789  \\
$\sigma_h$ & 41.489 $\pm$ 0.055 & 41.492 & 41.444  \\
$R_b$ & 0.2179 $\pm$ 0.0011 & 0.2156 & 0.2177  \\
$R_c$ & 0.1720 $\pm$ 0.0056 & 0.1720 & 0.1716  \\
$A_{FB}^e$ & 0.0161 $\pm$ 0.0025 & 0.0155 & 0.0170  \\
$A_{FB}^\mu$ & 0.0165 $\pm$ 0.0014 & 0.0155 & 0.0170  \\
$A_{FB}^\tau$ & 0.0204 $\pm$ 0.0018 & 0.0155 & 0.0170  \\
$A_{\tau}(P_\tau)$ & 0.1401 $\pm$ 0.0067 & 0.1440 & 0.1503  \\
$A_{e}(P_\tau)$ & 0.1382 $\pm$ 0.0076 & 0.1440 & 0.1503  \\
$A_{FB}^b$ & 0.0985 $\pm$ 0.0022 & 0.1010 & 0.0985  \\
$A_{FB}^c$ & 0.0734 $\pm$ 0.0048 & 0.0720 & 0.0755  \\
$A_{LR}$ & 0.1551 $\pm$ 0.0040 & 0.1440 & 0.1503  \\
$M_W$ & 80.38 $\pm$ 0.14 & 80.34 & 80.35  \\
$M_W/M_Z$ & 0.8814 $\pm$ 0.0011 & 0.8810 & 0.8811  \\
$g_L^2(\nu N \rightarrow \nu X)$ & 0.3003 $\pm$ 0.0039
      & 0.3030 & 0.3024  \\
$g_R^2(\nu N \rightarrow \nu X)$ & 0.0323 $\pm$ 0.0033
      & 0.0300 & 0.0297  \\
$g_{eA}(\nu e \rightarrow \nu e)$ & --0.503 $\pm$ 0.018
      & --0.507 & --0.506  \\
$g_{eV}(\nu e \rightarrow \nu e)$ & --0.025 $\pm$ 0.019
      & --0.037 & --0.039  \\
$Q_W(Cs)$ & --72.11 $\pm$ 0.93 & --72.88 & --72.56  \\
$R_{\mu \tau}$ & 0.9970 $\pm$ 0.0073 & 1.0000 & 1.0000  \\
\hline
\end{tabular}
\end{center}
\caption{ {\small Experimental
  \protect\cite{data}-\protect\cite{langacker}
  and predicted values of electroweak observables for the SM for
    $\alpha_s(M_Z)=0.115$. The SM values correspond to the
    best-fit values (with $m_t=173$ GeV, $m_{\rm Higgs} = 300$ GeV) in
    \protect\cite{langacker}, with $\alpha(M_Z)=1/128.9$,  and
    corrected for the change in $\alpha_s(M_Z)$.}}
\end{table}

  The best-fit values are
\vspace*{-2mm}
\bear
\delta g_L & = & 0.004 \pm 0.013~, \nonumber \\
\delta g_R^b & = & 0.036 \pm 0.068~, \nonumber \\
 S & = & -0.40 \pm 0.55~, \nonumber \\
 T & = & -0.25 \pm 0.46~.
\label{bestfit}
\eear
These values give a very good fit to the data; the $\chi^2$ (i.e. sum
of the squares of deviations over one standard-deviation errors) per degree of
freedom (df) is $\chi^2/{\rm df}  =  0.7$, while for the SM we find
$\chi^2/{\rm df}  =  1.50$. Thus we see that the SM (with $\alpha_s$
taken as an input rather than as an additional parameter to be
fit to the electroweak data \cite{bufits}) provides a
relatively poor fit to the data \cite{bufits}.  Another
way of stating this is that assuming the SM is correct,
the probability of  data giving a larger $\chi^2$ is only 6\%,
while assuming the extended model is correct, the probability of
finding a larger $\chi^2$ is 78\%.

For comparison, we also considered the case where
only the $S$ and $T$ parameters are added to the SM, and found that
the fit is almost as poor as in the case of the SM:
$\chi^2/{\rm df}  =  1.48$, corresponding to
$S  =  -0.09 \pm 0.34$ and $T  =  0.03 \pm 0.34$
(see also \cite{stnew}).

Note that in the absence of isospin violation effects other
than the one in eq.~(\ref{direct}),
the bound on $T$ provides a tight constraint on $\delta g_R^t$:
\be
\delta g_R^t < 0.02
\ee
at the 95\% confidence level (taking $\Lambda = 1$ TeV).  It is amusing that
measurements
below the top quark threshold can produce such a tight bound on the top
coupling.

%%%%%%%%%%%%%%%%%%%%%%%%%%%%%%%%%%%%%%%%%%%%%
%%%%%%%%%%%
%%%%%%%%%%%%%%%%%%%%%%%%%%%%%%%%%%%%%%%%%%%%%
%%%%%%%%%%%
\section{A technicolor model with weak singlet scalars}
\setcounter{equation}{0}

In technicolor models, the quark and lepton masses may be produced
by the exchange of gauge bosons \cite{etc}, weak-doublet scalars
\cite{tcsd}, or weak-singlet scalars \cite{kagan1}.
This latter alternative may explain certain features of the quark
and lepton spectrum, and does not require complicated dynamics.

As an example application for the effective field theory
calculation presented in Section 1, we examine in some detail
the effects of a technicolor model with weak-singlet scalars
on precision electroweak  measurements.
Because the main effects on electroweak observables are due to the
sector responsible for electroweak symmetry breaking and the $t$ and
$b$ masses, we will not be concerned with the mechanism
which generates the other fermion masses. Also, we will
assume the effects of the physics which keeps the scalars light
(e.g. compositeness or supersymmetry) to be negligible at a scale of order
1 TeV (note that the supersymmetry breaking
scale can be higher than in minimal supersymmetric extensions of the SM
since fine-tuning is not an issue).

In addition to the SM fermions, consider one doublet
of technifermions, \up\ and \down, and 3 scalars, $\phi$,
$\omega_t$, $\omega_b$,
which transform under the $SU(N_{\rm TC}) \times SU(3)_{\rm C}
\times SU(2)_{\rm W} \times U(1)_{\rm Y}$ gauge group
(with $N_{\rm TC}$ even) as:
\bear
& & \!\!\!\! \Psi_{\!\!\inr} =
\left(\!\!\ba{c}\up_{\inr} \\ \down_{\inr}\ea\!\!\right)
  \, : \; (N_{\rm TC}, 1, 2)_0 \; , \hspace{.5cm}
\up_{\inl} \, : \;  (N_{\rm TC}, 1, 1)_{+1}
\; , \hspace{.5cm}
\down_{\inl} \, : \;  (N_{\rm TC}, 1, 1)_{-1} \; ,
\nonumber \\ [4mm]
& & \!\!\!\! \phi \, : \; (N_{\rm TC}, \overline{3}, 1)_{- \frac{1}{3}}
\; , \hspace{.5cm}
\omega_t \, : \; (N_{\rm TC},  \overline{3}, 1)_{- \frac{7}{3}}
\; , \hspace{.5cm}
\omega_b \, : \; (N_{\rm TC},  \overline{3}, 1)_{\frac{5}{3}}
 \; ~.
\label{e1}
\eear

The most general Yukawa interactions are contained in
\be
{\cal L}_{\rm Y} = C_q \overline \Psi_{\!\!\inr} q_{\inl} \phi +
C_t \overline t_{\inr} \up_{\inl} \phi^{\dagger} +
C_b \overline b_{\inr} \down_{\inl} \phi^{\dagger} +
C_{\omega_t}  \overline t_{\inr} \down_{\inl} \omega_t^{\dagger} +
C_{\omega_b}  \overline b_{\inr} \up_{\inl} \omega_b^{\dagger} +
{\rm h.c.} ~,
\ee
where the Yukawa coupling constants, $C_q, C_t, C_b, C_{\omega_t},
C_{\omega_b}$, are defined to
be positive.

Assuming the $\phi$ techniscalar is sufficiently heavy to be
integrated out, this results in a $t$ mass
\be
m_t \approx \frac{C_qC_t}{M_\phi^2}\pi v^3
\left(\frac{3}{N_{\rm TC}} \right)^{\!1/2} ~.
\label{top}
\ee
The only four-fermion operators induced by techniscalar exchange
that will induce shifts in the electroweak couplings of the $t$ and
$b$ are
\bear
{\cal L}_{\rm eff}^{\rm 4F}=&& \!\!\!\!\!\!\!\!
- \left[\frac{C_t^2}{2 M_{\phi}^2}(\overline{\up}_{\inl}
\gamma^{\mu} \up_{\inl}) + \frac{C_{\omega_t}^2}{2 M_{\omega_t}^2}
(\overline{\down}_{\inl} \gamma^{\mu} \down_{\inl}) \right]
(\overline{t}_{\inr}\gamma_{\mu} t_{\inr})
- \frac{C_{\omega_b}^2}{2 M_{\omega_b}^2} (\overline{\up}_{\inl}
\gamma^{\mu} \up_{\inl}) (\overline{b}_{\inr}\gamma_{\mu} b_{\inr})
\nonumber \\ [2mm]
&& \!\!\!\!\!\!\!\!
- \frac{C_q^2}{4 M_\phi^2}
\left[ \left(\overline \Psi_{\!\!\inr} \gamma^\mu \sigma_j
\Psi_{\!\!\inr} \right)
\left(\overline q_{\inl} \gamma_\mu  \sigma_j q_{\inl}  \right)
+ \left(\overline \Psi_{\!\!\inr} \gamma^\mu \Psi_{\!\!\inr} \right)
\left(\overline q_{\inl} \gamma_\mu q_{\inl} \right) \right] ~.
\label{ops}
\eear

These operators have an effect on the $Z$
couplings to $t \overline{t}$ and $b \overline{b}$ which can be
evaluated using an effective Lagrangian approach \cite{css}:
\bear
\overline \Psi_{\!\!\inr} \gamma^\mu \sigma_j
\Psi_{\!\!\inr}
& = & - i \frac{v^2}{2}
{\rm Tr} \left(D^{\mu} \, \Sigma \sigma_j
\Sigma^{\dagger} \right) ~,
\nonumber\\ [2mm]
\overline{\up}_{\inl} \gamma^{\mu} \up_{\inl} & = & i \frac{v^2}{2}
{\rm Tr} \left(\Sigma^{\dagger} \frac{\sigma_3+1}{2} D^{\mu} \Sigma
\right) ~,
\nonumber\\ [2mm]
\overline{\down}_{\inl} \gamma^{\mu} \down_{\inl} & = &  i
\frac{v^2}{2}
{\rm Tr} \left(\Sigma^{\dagger} \frac{-\sigma_3+1}{2} D^{\mu} \Sigma
\right) ~.
\eear
Since $\Sigma$ transforms as $C \Sigma W^{\dagger}$ under
$SU(2)_C \times SU(2)_W$ (where $SU(2)_C$ is the custodial symmetry
which has $U(1)_Y$ as subgroup), the covariant derivative is given by
\be
D^{\mu} \Sigma = \partial^{\mu} \Sigma +
i g \Sigma \frac{\sigma^k}{2} W^{\mu}_k - i g^\prime
\frac{\sigma^3}{2} B^{\mu} \Sigma ~.
\ee
With these expressions for operators, a comparison of
eqs.~(\ref{ops}) and (\ref{vertex}) yields
the couplings induced by scalar exchange [we eliminate $M_\phi$ using
eq.~(\ref{top})]:
\bear
\delta g_{L} & = & - \frac{C_q}{C_t} \frac{m_t}{8 \pi v}
\left(\frac{N_{\rm TC}}{3} \right)^{\!1/2} ~,
\nonumber
\\ [2mm]
\delta g_R^{t} & = & \frac{C_t}{C_q} \frac{m_t}{8 \pi v}
\left(\frac{N_{\rm TC}}{3} \right)^{\!1/2}
- \frac{C_{\omega_t}^2 v^2}{8 M_{\omega_t}^2} ~,
\nonumber
\\ [2mm]
\delta g_R^{b} & = & \frac{C_{\omega_b}^2 v^2}{8 M_{\omega_b}^2} ~.
\label{deltas}
\eear

The technifermion contribution to $S$ is estimated to be \cite{ST}
\be
S^0 \approx 0.1 \ N_{\rm TC} ~,
\label{stc}
\ee
so that, for\footnote{It may be possible to have $N_{\rm TC} = 2$ if the
scalars have a
significant effect on the vacuum alignment \cite{vacuum}, this issue has not
been studied in detail in the
literature.} $N_{\rm TC} = 4$ and $\Lambda \approx 1$ TeV, the
prediction of this model is
\be
S = S^0 + S^{(t,b)}
\approx  0.4 - 1.02 \left( 3\, \delta g_L - 2\, \delta g_{R}^{t}
+ \delta g_{R}^{b} \right) ~.
\label{Sfinal}
\ee

In addition to the direct isospin violation (\ref{direct}),
there are ``indirect''
contributions to $T$ from the technifermion mass spectrum
which can be only roughly estimated \cite{cdt2}:
\be
T^0  \sim \frac{N_{\rm TC}}{16 \pi^2 \alpha v^2}
\left( \Sigma_{\up}(0) - \Sigma_{\down}(0) \right)^2  ~,
\ee
where $\Sigma_{\up}(q^2)$ and $\Sigma_{\down}(q^2)$  are the
technifermion self-energies. In this model, the origin of the
indirect isospin violation is the difference between
the $C_t$ and $C_b$ Yukawa couplings which accounts for the $t-b$
mass splitting. The four-fermion operator responsible for $m_t$,
\be
\frac{C_tC_q}{M_\phi^2} \left(
\overline{\Psi}_{\inr} q_{\inl}\right) (\overline{t}_{\inr}
\up_{\inl}) ~,
\ee
gives a one-loop correction to $\Sigma_{\up}$ which is quadratic
divergent:
\be
\Sigma_{\up}(0) - \Sigma_{\down}(0) = - \frac{3}{16 \pi^3} \,
\frac{m_t^2}{v^3} \Lambda^{\prime 2} ~,
\ee
where $\Lambda^\prime$ is expected to be of order 1 TeV, and
generically different than $\Lambda$.
Thus, the result is extremely sensitive to $\Lambda^\prime$,
\be
T^0  \approx 0.51 \,
\left(\frac{N_{\rm TC}}{4}\right)
\left(\frac{\Lambda^\prime}{\rm 1 TeV}\right)^{\! 4} ~,
\ee
and it is not possible to evaluate precisely this isospin breaking
effect. For $N_{\rm TC} = 4$, the final expression for $T$ is
\be
T = T^0 + T^{(t,b)}
\approx 0.51 \, \left(\frac{\Lambda^\prime}{\rm 1 TeV}\right)^{\!
  4} + 34.0 \, \delta g_{R}^{t} ~.
\label{Tfinal}
\ee

It is interesting that all the contributions from $\omega_t$ and
$\omega_b$ exchange decrease $S$ and $T$, as can be seen from
eqs.~(\ref{deltas}), (\ref{Sfinal}) and (\ref{Tfinal}).
For a range of values of the parameters $C_{t,q,\omega_t,\omega_b}, \,
M_{\omega_t,\omega_b}$ and $\Lambda^\prime$,
the predictions for $\delta g_L, \, \delta g_{R}^{b}, \, S$ and $T$,
are within the correct ballpark (the best-fit values of eq. (\ref{bestfit})).
For example, for
$C_t/C_q = 2.5$, $C_{\omega_t} (0.5 \, {\rm TeV})/M_{\omega_t} = 2$,
$C_{\omega_b} (0.5 \, {\rm TeV})/M_{\omega_b} = 1.7$ and
$\Lambda^\prime = 1.2$ TeV, we find
\be
\delta g_L \approx - 0.013 ~, \; \delta g_{R}^{b} \approx 0.088 ~,
\; S \approx 0.27 ~, \; T \approx - 0.32 ~.
\ee
We emphasize, though, that given the sensitivity of $T^0$ to
$\Lambda^\prime$,
and the fact that the value of $\Lambda^\prime$ is fixed by the
dynamics and not known precisely, the prediction for $T$ is only
potentially successful.

%%%%%%%%%%%%%%%%%%%%%%%%%%%%%%%%%%%%%%%%%
%%%%%%%%%%%%%%%%%%%%%%
\section{Conclusions}

As we have seen in Section 3, if the SM were the correct theory of
nature, then the probability of producing such
a poor fit to electroweak measurements is only 6\%. Therefore, it is
worthwhile
to look for extensions or alternatives to the SM which provide better
fits to the
existing data. The effective theory discussed in this paper greatly
improves the fit (the $\chi^2/{\rm df}$ is half as large)
by including four additional parameters:
two in the gauge boson self-energies and two in the gauge boson
couplings to third generation quarks.

It is interesting that this effective theory can arise below the
weak scale from a technicolor model which offers insight into the
pattern of quark and lepton masses (for example, the mass hierarchy
between the three generations may be induced by the masses of the
exchanged scalars \cite{kagan1}).
Given that compositeness and/or supersymmetry seem to be required to
keep the scalars light, this technicolor model can only be considered
as an effective theory
valid up to roughly 1 TeV. However, to decide whether this
model is a better candidate than the SM, one would need improved
methods of computing the isospin breaking effects.

It should be stressed that the class of high energy theories
which may lead to the effective theory discussed in Sections 2 and 3
is potentially large. For example, the analysis presented in Section 4
can be extended to any strong dynamics which gives rise
to four-fermion interactions. Of course, the sizes and signs of the
contributions to the $S$ and $T$ parameters are model dependent.

%%%%%%%%%%%%%%%%%%%%%%%%%%%%%%%%%%%%%%%%%
%%%%%%%%%%%%%%%%%%%%%%
\vspace{12pt} \centerline{\bf Acknowledgments} \vspace{2mm}

We would like to thank Sekhar Chivukula for helpful discussions and
comments on the manuscript.
{\em This work was supported in part by the National Science
  Foundation under grants PHY-9057173, and PHY-95-14797;
and by the Department of Energy under grant DE-FG02-91ER40676.}

%%%%%%%%%%%%%%%%%%%%%%%%%%%%%%%%%%%%%%%%%
%%%%%%%%%%%%%%%%%%%%%%
\appendix
\section*{Appendix}
\renewcommand{\theequation}{A.\arabic{equation}}
\setcounter{equation}{0}

In this Appendix we list the deviations from the SM predictions for
the electroweak data in terms of $\delta g_L, \, \delta g_R^b, \, S$
and $T$.
\begin{equation}
\Gamma_Z = \left( \Gamma_Z \right)_{SM} \left(1 -3.8 \times 10^{-3} S
 + 0.010 \, T -0.707 \, \delta g_L + 0.128 \, \delta g_R^b \right)
\end{equation}
\begin{equation}
 R_{e,\mu,\tau} = \left( R_{e,\mu,\tau} \right)_{SM} \left(1 -2.9
 \times 10^{-3} S + 2.0 \times
10^{-3} T -1.01 \,  \delta g_L + 0.183 \,  \delta g_R^b \right)
\end{equation}
\begin{equation}
 \sigma_h = \left( \sigma_h \right)_{SM} \left(1 + 2.2 \times 10^{-4}
 S -1.6
\times 10^{-4} T + 0.404 \,  \delta g_L -0.073 \,  \delta g_R^b
\right)
\end{equation}
\begin{equation}
 R_b = \left( R_b \right)_{SM} \left(1 + 6.6 \times 10^{-4} S -4.0
 \times
10^{-4} T -3.56 \,  \delta g_L + 0.645 \,  \delta g_R^b \right)
\end{equation}
\begin{equation}
 R_c = \left( R_c \right)_{SM} \left(1 -1.3 \times 10^{-3} S + 10.0
 \times 10^{-4} T + 1.01 \,  \delta g_L -0.183 \,  \delta g_R^b
\right)
\end{equation}
\begin{equation}
 A_{FB}^{e,\mu,\tau} = \left( A_{FB}^{e,\mu,\tau} \right)_{SM} -6.8
 \times 10^{-3} S + 4.8 \times 10^{-3} T
\end{equation}
\begin{equation}
 A_{\tau}(P_\tau) = \left( A_{\tau}(P_\tau) \right)_{SM} -0.028 \, S
  + 0.020 \,  T
\end{equation}
\begin{equation}
 A_{e}(P_\tau) = \left( A_{e}(P_\tau) \right)_{SM} -0.028 \,  S +
 0.020 \, T
\end{equation}
\begin{equation}
 A_{FB}^b = \left( A_{FB}^b \right)_{SM} -0.020 \,  S + 0.014 \,  T
 -0.035 \,  \delta g_L -0.191 \,  \delta g_R^b
\end{equation}
\begin{equation}
 A_{FB}^c = \left( A_{FB}^c \right)_{SM} -0.016 \,  S + 0.011 \,   T
\end{equation}
\begin{equation}
 A_{LR} = \left( A_{LR} \right)_{SM} -0.028 \,  S + 0.020 \,  T
\end{equation}
\begin{equation}
 M_W = \left( M_W \right)_{SM} \left(1 -3.6 \times 10^{-3} S + 5.5
 \times 10^{-3} T \right)
\end{equation}
\begin{equation}
 M_W/M_Z = \left( M_W/M_Z \right)_{SM} \left(1 -3.6 \times 10^{-3} S +
 5.5 \times 10^{-3} T \right)
\end{equation}
\begin{equation}
 g_L^2(\nu N \rightarrow \nu X) = \left( g_L^2(\nu N \rightarrow \nu
 X) \right)_{SM} -2.7 \times 10^{-3} S + 6.6 \times 10^{-3} T
\end{equation}
\begin{equation}
 g_R^2(\nu N \rightarrow \nu X) = \left( g_R^2(\nu N \rightarrow \nu
 X) \right)_{SM} + 9.4 \times 10^{-4} S -1.9 \times 10^{-4} T
\end{equation}
\begin{equation}
 g_{eA}(\nu e \rightarrow \nu e) = \left( g_{eA}(\nu e \rightarrow \nu
 e) \right)_{SM} -3.9 \times 10^{-3} T
\end{equation}
\begin{equation}
 g_{eV}(\nu e \rightarrow \nu e) = \left( g_{eV}(\nu e \rightarrow \nu
 e) \right)_{SM} + 7.2 \times 10^{-3} S -5.4 \times 10^{-3} T
\end{equation}
\begin{equation}
 Q_W(Cs) = \left( Q_W(Cs) \right)_{SM} -0.796 \,  S -0.011 \,  T
\end{equation}
\begin{equation}
 R_{\mu \tau} = \left( R_{\mu \tau} \right)_{SM}
\end{equation}

%%%%%%%%%%%%%%%%%%%%%%%%%%%%%%%%%%%%%%%%%
%%%%%%%%%%%%%%%%%%%%%%
\newcommand{\np}{{\it Nucl.\ Phys.}\ {\bf B}}
\newcommand{\pr}{{\it Phys.\ Rev.}\ }
\newcommand{\prd}{{\it Phys.\ Rev.}\ {\bf D}}
\newcommand{\prp}{{\it Phys.\ Rep.}\ }
\newcommand{\prl}{{\it Phys.\ Rev.\ Lett.}\ }
\newcommand{\pl}{{\it Phys.\ Lett.}\ {\bf B}}
\newcommand{\ptp}{{\it Prog.\ Theor.\ Phys.}\ }
\newcommand{\ap}{{\it Ann.\ Phys.}\ }
\newcommand{\intl}{{\it Int.\ J.\ Mod.\ Phys.}\ {\bf A}}

\vfil
\end{document}